\pdfoutput=1
\documentclass[aps,groupedaddress,tightenlines,11pt]{revtex4-2}

\usepackage{xspace}

\usepackage[utf8]{inputenc}
\usepackage[T1]{fontenc}
\usepackage{csquotes}
\usepackage[english]{babel}
\usepackage{graphicx}
\graphicspath{{Pictures/}{/}}
\usepackage[export]{adjustbox}	

\usepackage[dvipsnames]{xcolor}
\usepackage{tikz-cd}
\usepackage{array}

\usepackage[final]{fixme}
\usepackage[nointlimits]{amsmath}
\usepackage{amssymb,mathrsfs,mathtools}

\usepackage{upgreek}
\usepackage{braket}
\usepackage{cancel}
\usepackage{siunitx}
\usepackage{aliascnt} 
\usepackage[amsmath,thmmarks]{ntheorem}
\usepackage[expansion=true,protrusion=true]{microtype}
\usepackage{xkeyval}
\usepackage{overpic}

\usepackage[final,%
	pdftex,%
	bookmarks,%
	bookmarksdepth=3,%
	breaklinks=true,%
	colorlinks=true,%
	urlcolor=NavyBlue,%
	linkcolor=NavyBlue,%
	citecolor=ForestGreen,%
]{hyperref}%
\usepackage[all]{hypcap}

\DeclarePairedDelimiterXPP{\pars}[1]{\mathop{}}{\lparen}{\rparen}{}{#1}

\DeclarePairedDelimiterXPP{\abs}[1]{\mathop{}}{\lvert}{\rvert}{}{#1}
\DeclarePairedDelimiterXPP{\norm}[1]{\mathop{}}{\lVert}{\rVert}{}{#1}
\DeclarePairedDelimiterXPP{\seminorm}[1]{\mathop{}}{\lbrack}{\rbrack}{}{#1}
\DeclarePairedDelimiterXPP{\inner}[1]{\mathop{}}{\langle}{\rangle}{}{#1}
\DeclarePairedDelimiterXPP{\brackets}[1]{\mathop{}}{\lbrack}{\rbrack}{}{#1}
\DeclarePairedDelimiterXPP{\braces}[1]{\mathop{}}{\lbrace}{\rbrace}{}{#1}

\DeclarePairedDelimiterXPP{\intervalcc}[1]{\mathop{}}{\lbrack}{\rbrack}{}{#1}
\DeclarePairedDelimiterXPP{\intervalco}[1]{\mathop{}}{\lbrack}{\rparen}{}{#1}
\DeclarePairedDelimiterXPP{\intervaloc}[1]{\mathop{}}{\lparen}{\rbrack}{}{#1}
\DeclarePairedDelimiterXPP{\intervaloo}[1]{\mathop{}}{\lparen}{\rparen}{}{#1}

\DeclarePairedDelimiterXPP{\myset}[2]{\mathop{}}{\lbrace}{\rbrace}{}{#1\,\delimsize\vert\,\mathopen{}#2}

\let\dot\undefined

\DeclarePairedDelimiterXPP{\dot}[2]{\mathop{}}{\langle}{\rangle}{}{#1,#2}

\DeclarePairedDelimiterXPP{\floor}[1]{\mathop{}}{\lfloor}{\rfloor}{}{#1}
\DeclarePairedDelimiterXPP{\ceil}[1]{\mathop{}}{\lceil}{\rceil}{}{#1}

\DeclareDocumentCommand{\converges}{ o }{
	\mathbin{%
		\IfValueTF{#1}{%
			\mathrel{\vbox{\offinterlineskip\ialign{%
				\hfil##\hfil\cr
				$\scriptscriptstyle#1$\cr
				$-\!\!\!-\!\!\!\rightarrow$\cr
			}}}
		}{%
			-\!\!\!-\!\!\!\rightarrow
		}%
	}%
}

\newcommand{\numtoword}[1]{%
  \ifcase#1\relax 
    zero
  \or one
  \or two
  \or three
  \or four
  \or five
  \or six
  \or seven
  \or eight
  \or nine
  \or ten
  \or eleven
  \or twelve
  \or thirteen
  \or fourteen
  \or fifteen
  \or sixteen
  \or seventeen
  \or eighteen
  \or nineteen
  \or twenty
  \or twentyone
  \or twentytwo
  \or twentythree
  \or twentyfour
  \or twentyfive
  \or twentysix
  \or twentyseven
  \else ERROR
  \fi
}

\newcommand{\computedlambda}[2]{%
  \csname computedlambda\numtoword{#1}\numtoword{#2}\endcsname
}

\newcommand{\computedC}[2]{%
\csname computedC\numtoword{#1}\numtoword{#2}\endcsname
}

\newcommand{\totalPolygonsSampled}{1.72\times 10^{10}}
\newcommand{\totalSummands}{2.5\times 10^8}

\newcommand{\totalSummandsNotIdentifiedOnFirstPass}{17}

\usepackage[capitalise,nameinlink]{cleveref}

\theoremstyle{break}

\theoremstyle{plain}
\theorembodyfont{\normalfont}

{%
    \theorembodyfont{\footnotesize}
    \theoremsymbol{\ensuremath{\Diamond}}%
}

\theoremstyle{break}

\theoremheaderfont{\itshape}
\theorembodyfont{\upshape}
\theoremstyle{nonumberplain}
\theoremseparator{.}
\theoremsymbol{\ensuremath{\Box}}

\begin{document}

\title{Random knotting in very long off-lattice self-avoiding polygons}

\author{Jason Cantarella}
\affiliation{Mathematics Department, University of Georgia, Athens, GA, USA}
\author{Tetsuo Deguchi}
\affiliation{Department of Physics, Ochanomizu University, Bunkyo-ku, Tokyo, Japan}
\author{Henrik Schumacher}
\affiliation{Institut für Mathematik, RWTH Aachen University, Aachen, Germany}
\author{Clayton Shonkwiler}
\affiliation{Department of Mathematics, Colorado State University, Fort Collins, CO, USA}
\author{Erica Uehara}
\affiliation{Graduate School of Informatics, Kyoto University, Sakyo-ku, Kyoto, Japan}

\date{\today}

\begin{abstract}
		We present experimental results on knotting in off-lattice self-avoiding polygons in the bead-chain model. Using Clisby's tree data structure and the scale-free pivot algorithm, for each $k$ between $10$ and $27$ we generated $2^{43-k}$ polygons of size $n=2^k$. Using a new knot diagram simplification and invariant-free knot classification code, we were able to determine the precise knot type of each polygon. The results show that the number of prime summands of knot type $K$ in a random $n$-gon is very well described by a Poisson distribution. We estimate the characteristic length of knotting as $\num{656500} \pm \num{2500}$. We use the count of summands for large $n$ to measure knotting rates and amplitude ratios of knot probabilities more accurately than previous experiments. Our calculations agree quite well with previous on-lattice computations, and support both knot localization and the knot entropy conjecture.  
		
\end{abstract}

\maketitle


\section{Introduction}

Self-avoiding polygons (SAPs) with knotted topologies have attracted much interest in various branches of polymer physics, chemistry, and biology because they provide a model for knotted ring polymers in nature and also in synthetic chemistry. 
For example, these polymers have been found in living organisms such as the circular DNA of \emph{E. coli} \cite{Nature1983_KrasnowCozzarelli,PNAS1993_RybenkovVologodskii,Science1993_ShawWang,NatCommun2019_SharmaGaraj}. 

As suggested by Edwards~\cite{JPhysA1968_Edwards}, it is nontrivial to study the statistical properties of a ring polymer in solution or melt. Such a polymer does not change its topology under thermal fluctuations since it cannot pass through itself. This topological constraint leads to a large reduction in the available volume or degrees of freedom in the configuration space of the ring polymer and hence affects the statistical properties of the ring polymer(s). For example, knotted ring polymer melts have a different crystallization behavior than unknotted ring polymer melts \cite{macromolecules2023_HagitaFujiwara}. 
The reticulated structure of synthesized polymer networks may have self-entanglements, and it is suggested that they should enhance the elastic modulus \cite{JChemPhys1977_GraessleyPearson}.

In the 1960s, Frisch, Wasserman and Delbr\"{u}ck~\cite{frisch1961,ProcSympApplMath1962_DelbrueckFueller} addressed the conjecture that nontrivial knotted ring polymers should be very common in large ring polymers. Knotting probability has been evaluated in numerical simulations by making use of knot invariants \cite{JETP1974_VologodskiiAnshelevich,JPhysFrance1979_CloizeauxMehta,Biopolymers1980_LeBret,
PhysLettA1982_MichelsWiegel,KoniarisMuthukumar1991,JKTR1994_DeguchiTsurusaki,PhysRevE1997_DeguchiTsurusaki,UeharaKnotting2017,JPhysA2021_XiongWhittington,JPhysA2024_CantarellaShonkwiler} and measured directly in biological experiments of DNA~
\cite{PNAS1993_RybenkovVologodskii,Science1993_ShawWang}. Furthermore, it was rigorously shown that a self-avoiding walk is exponentially likely to be knotted if it is very long \cite{JPhysA1988_SumnersWhittington,DiscreteApplMath1989_Pippenger}. In all of these contexts, the flexibility and thickness of the polymer plays an important role. 

The asymptotic knotting behavior of long and thick self-avoiding polygons (SAPs) with fixed topology is still not fully understood. Knots are generally considered localized and asymptotically independent events on a long-enough polymer~\cite{PhysRevE2000_Katritch_Stasiak,JPhysA2005_Marcone,PhysRevE2007_Marcone,PhysBiol2009_OrlandiniVanderzande}, and there is strong numerical evidence for this idea~\cite{MR99f:82065}, at least for relatively short polygons with a relatively small number of summands. In this paper, we consider the random variable
$$
\begin{aligned}
m_K^n &= \text{number of prime summands of knot type $K$ in an $n$-gon.}
\end{aligned}
$$
If prime summands are independent and localized, then $m_K^n$ should be approximately Poisson-distributed; that is, we should have 
\begin{equation}
P(m_K^n = m) \approx \frac{(\lambda_K(n))^m e^{-\lambda_K(n)}}{m!}, 
\label{eq:fundamental equation}
\end{equation}
where $\lambda_K(n)$ is the expected value of $m_K^n$~\footnote{For very large $m$, we know that $P(m_K^n = m) = 0$, while $\frac{(\lambda_K(n))^m e^{-\lambda_K(n)}}{m!} > 0$, so the two sides of~\eqref{eq:fundamental equation} cannot be exactly equal.}. This is an old observation in the random knotting community, probably dating back to the 1960's~\cite{WhittingtonPC}.

Assume that we are in a range of $n$, $m$, and $K$ where~\eqref{eq:fundamental equation} is a good approximation. Let $R_K(n) = \lambda_K(n)/n$ be the rate of production of summands of knot type $K$ (per edge) in an $n$-gon. It would follow that the probability of finding a polygon with no prime summands of type $K$ would be well-approximated by
$$
P(m_K^n = 0) \approx \frac{(R_K(n) n)^0 e^{-R_K(n) n}}{0!} = e^{-R_K(n) n}. 
$$
Assuming that $m_K^n$ and $m_{K'}^n$ are approximately independent when $K$ and $K'$ are distinct prime knot types, this immediately implies that the probability of an unknot
\begin{equation}
\begin{aligned}
P_{0_1}(n) &\approx \Pi_{K \text{a prime knot type}}\, P(m_K^n = 0) \\
&= \Pi_{K \text{a prime knot type}}\, e^{-R_K(n) n} \\
&= e^{-\left(\sum_K R_K(n)\right) n},
\end{aligned}
\label{eq:probability of unknot}
\end{equation}
while the probability $P_K(n)$ that an $n$-gon has (prime) knot type $K$ is
\begin{equation}
\begin{aligned}
P_K(n) &= P(m_K^n = 1) \Pi_{K' \neq K} P(m_{K'}(n) = 0) \\
       &\approx R_K(n) \, n \, e^{-R_K(n) n}\, e^{-\sum_{K' \neq K} R_K'(n) n} \\
       &= R_K(n)\, n\, e^{-\left(\sum_{K'} R_{K'}(n) \right) n} \\
       &= R_K(n) \, n \, P_{0_1}(n).
\end{aligned}
\label{eq:knot probability with finite size effect model}
\end{equation}
 
More generally, the \emph{knot entropy conjecture}~\cite{beatonFirstProofKnot2024,beatonEntanglementStatisticsPolymers2026} (cf.~\cite{orlandiniEntropicExponentsLattice1996,orlandiniAsymptoticsKnottedLattice1998}) says that for any (prime or composite) knot type $\overline{K}$ the probability $P_{\overline{K}}
(n)$ is well-approximated by
\begin{equation}
P_{\overline{K}}(n) \approx C_{\overline{K}} n^{m(\overline{K})} P_{0_1}(n) \left(1 + \frac{\beta_{\overline{K}}}{n^{\Delta}} + \frac{\gamma_{\overline{K}}}{n} \right) ,
\label{eq:standard model}
\end{equation}
where $m(\overline{K})$ is the number of prime factors of $\overline{K}$, we say $m(0_1) = 0$, and
\begin{equation}
P_{0_1}(n) \approx A_{0_1} n^{\alpha_{0_1}} \mu_{0_1}^n.
\label{eq:unknot probability}
\end{equation}
Here, the parameters $\beta_{\overline{K}}$ and $\gamma_{\overline{K}}$ describe the finite-size correction. The parameter $\alpha_{0_1}$ (the unknot critical exponent) and the dimensionless ratios $C_{\overline{K}}/C_{\overline{K}'}$ should be universal, while the other parameters $C_{\overline{K}}$, $\Delta$, $A_{0_1}$, and $\mu_{0_1}$ are model-dependent. 

The characteristic length of the unknot is given by $N_{0_1}:=-1/\log{\mu_{0_1}}$ and the unknotting probability is approximately  $\exp(-n/N_{0_1})$. If the number of segments $n$ is much larger than $N_{0_1}$, this implies that the polygon is very likely to be knotted.

Comparing~\eqref{eq:probability of unknot} and~\eqref{eq:unknot probability}, we see that $A_{0_1} = 1$ and $\alpha_{0_1} = 0$ in this model, while
\begin{equation}
\mu_{0_1}=e^{-\left(\sum_K R_K(n)\right)}.
\end{equation}
Since $m_K^n = 1$ for a prime knot, combining \eqref{eq:knot probability with finite size effect model}, \eqref{eq:standard model}, and~\eqref{eq:unknot probability} yields
\begin{equation}
R_K(n) \approx C_K \left( 1 + \frac{\beta_K}{n^{\Delta}} + \frac{\gamma_K}{n} \right).
\label{eq:rate with finite size effect model}
\end{equation}
 
It is difficult to test~\eqref{eq:standard model} directly because $P_{K}(n) \rightarrow 0$ exponentially fast as $n \rightarrow \infty$ for any prime knot $K$, making it difficult to gather data for large $n$~(see \cite{JANSEVANRENSBURG:2011jo,JPhysCondensMatter2015_UeharaDeguchi}). This has made it very hard to compute the values of the $C_K$ accurately. However, we can compute the $C_K$ and therefore estimate $P_K(n)$ by gathering data on $R_K(n)$. Further, as noted by Janse van Rensburg~\cite{jansevanrensburgProbabilityKnottingLattice2002},
\[
\frac{C_K}{C_{K'}} = \lim_{n \rightarrow \infty} \frac{P_K(n)}{P_{K'}(n)} = \lim_{n \rightarrow \infty} \frac{R_K(n)}{R_{K'}(n)} = \lim_{n \rightarrow \infty} \frac{\lambda_K(n)}{\lambda_{K'}(n)}.
\]
That is, the amplitude ratio is the ratio of the mean number of prime summands of knot type $K$ to the mean number of prime summands of knot type $K'$ in a random self-avoiding $n$-gon. While the $C_K$ values are model-dependent, we expect that this ratio should be universal~(see~\autoref{tab:amplitude ratios}).

In this paper, we gather experimental data to test the approximations above in the off-lattice ``string of pearls'' model~\footnote{Here, a ring polymer is modeled by interior-disjoint unit diameter spheres centered at $v_1, \dots, v_n$ where the spheres centered at $v_i$ and $v_{i+1}$ (as well as $v_n$ and $v_1$) must be tangent~\cite{StellmanGans1972,KenyonWinkler2009,McMullen2018}.} for self-avoiding polygons. We focus on the medium-to-large $n$ regime, using new computational tools which enable us to generate extremely large polygons and classify their knot types exactly. 

We perform four experimental tests. First, we compute the empirical distribution of $m_K^n$ for knot types through $6$ crossings and $n = 2^k$ with $k \in \{10,11,\dotsc,27\}$. For $2^k$-gons, we sampled at least $2^{43-k}$ polygons, so the total number of edges sampled was at least $2^{43}\approx 8.8\times10^{12}$ for each $k$. We check that the empirical distribution is very closely approximated by a Poisson distribution with the corresponding empirical mean $\lambda_K(n)$. These empirical means become quite large and are hence easy to sample accurately for large $n$. For example, $\lambda_{3_1}({2^{27}}) \approx \computedlambda{1}{27}$ (the uncertainly is standard error). 

Second, we compute knotting rates $R_K(n)$ for the same $K$ and $n$ and find that~\eqref{eq:rate with finite size effect model} explains the finite-size effects very well.

Third, we use these knotting rates to estimate the knotting probability for the unknot and compare the results with counts of unknots in our data. We see a good match to the ``pure exponential'' model given by~\eqref{eq:unknot probability}, and we are able to  estimate the characteristic length of knotting ($\num{656500} \pm \num{2500}$).

Last, we return to~\eqref{eq:standard model}, and compute knotting probabilities $P_K(n)$ for simple knots, finding that~\eqref{eq:knot probability with finite size effect model} fits well and yields almost exactly the same value of $C_K$ (and comparable values of $\beta_K$ and $\gamma_K$) for trefoils. The remaining knot types show significant differences in all three parameters. We take this as an indication of the difficulty in computing $C_K$ from $P_K(n)$ data.

Overall, our results are fully consistent with the hypothesis that the Poisson model~\eqref{eq:fundamental equation} is a useful description of the knotting of random self-avoiding polygons. They support knot localization (and hence the knot entropy conjecture) in a range of $n$ much larger than previous experiments have been able to test. Further, the standard model of the finite-size correction~\eqref{eq:knot probability with finite size effect model} is also supported by our data. 
 
\section{Methodology}

To generate polygons, we used a modification of the classical polygonal fold method~\cite{MadrasSokalPivot}. In each trial step, a vertex $v_{i}$ is chosen uniformly on the self-avoiding polygon. For an $n$-gon, we then sample $x$ uniformly on $[1,\log_2(n)]$, and round $2^x$ to the nearest integer $k$, as suggested by the ``scale-free sampling'' procedure of Clisby~ \cite{clisbyCalculationConnectiveConstant2013}. The ``mobile arc'' of the polygon is then defined to be the edges between vertices $v_{i}$ and $v_{i \pm k}$ (with equal probability). The proposed move is uniformly sampled from the rotations around the line through $v_{i}$ and $v_{i \pm k}$ and the reflections over planes containing $v_{i}$ and $v_{i \pm k}$. In every case, a proposed step is accepted if the new configuration is still self-avoiding. We measured acceptance probability $\approx 0.63 \, n^{-0.057}$. This algorithm has detailed balance, so it samples uniformly. We have not proved that every bead-chain SAP is accessible from our starting configuration (the regular $n$-gon), though we conjecture that this is true.

We burned in each run for $20\,n$ attempted steps and then took samples at intervals of $n/30$ attempted steps. We used 64 parallel Markov chains with these parameters for all $n$ except $2^{27}$, where we took $128$ parallel chains. We wrote a new, highly-performant implementation of Clisby's tree~\cite{ClisbyEfficient2010,ClisbyAccurate2010} (see also \cite{clisbyOfflatticeParallelImplementations2021, SchnabelJanke2023, SchnabelJanke2020}), a data structure which provides $O(\log n)$ folds (including collision checking) while retaining $O(n)$ computation of all vertex positions. Our implementation requires considerable memory for very large polygons ($\approx 30$ gb for $2^{27}$-gons), so we ran $2^{24}$–$2^{27}$-gons on the high-memory nodes of System C (Cinnamon) at Kyoto University and the remaining calculations on the Sapelo2 cluster at the Georgia Advanced Computing Resource Center (GACRC) at the University of Georgia. Our code~\texttt{polyfold} is available on GitHub~\cite{schumacherKnoodle}.

The self-avoiding polygons we generated had a large number of crossings in their planar projections. On average, projections of $2^{27}$-gons had $\approx 2^{25}$ crossings.  We constructed planar diagram codes for these knots with verified accuracy using extremely careful floating point computation for segment intersections~(see \cite{Boldo2009, Jeannerod2013}). To identify the resulting knots we wrote a new code called~\texttt{Knoodle}~\cite{schumacherKnoodle,cantarellaHardUnknotsAre} for very efficient combinatorial simplification of knot diagrams. The key idea is that an arc of a diagram that only crosses over other strands may be rerouted as desired to reduce the number of crossings in the diagram (this is called a ``pass move'' in knot theory, and we call this process ``pass reduction'').
\texttt{SnapPy} will perform similar simplifications, but requires (at least) $O(n)$ memory and time to find each move, so the entire simplification is $O(n^2)$. Our code finds simplifications in essentially constant time and memory, making the entire process roughly linear in time (see~\cite{cantarellaHardUnknotsAre}). In practice, it is several orders of magnitude faster than \texttt{SnapPy}.

Further, portions of a knot diagram connected by only two arcs to the rest may be identified as connect summands and isolated immediately for further simplification. This process almost always results in a topologically minimal crossing diagram. When it does not, we use graph embedding techniques to construct a new lattice curve with the same knot type as the diagram, and rotate it to a different view, then resume the pass reduction process (again, see~\cite{cantarellaHardUnknotsAre} for more detail). \texttt{Knoodle} is available on GitHub~\cite{schumacherKnoodle}.

In total, we identified $\approx \totalSummands$ prime summands in our $\approx \totalPolygonsSampled$ polygons. All but $\totalSummandsNotIdentifiedOnFirstPass$ had $\leq 13$ crossings after our simplification process. We were able to classify these very rapidly without using knot invariants by matching them to a precomputed list of diagrams~(see \cite{cantarellaVeryFastCode}). The remaining $\totalSummandsNotIdentifiedOnFirstPass$ knots were all hyperbolic, and we determined their knot types by recognizing their knot complements using~\texttt{SnapPy}.

Although we were able to distinguish symmetry types of knots (such as $3_1$ and its mirror image, $3_1^m$) in our identifications of knot types, we grouped them together in our analysis to match the existing literature (for example writing $3_1/3_1^m$ to indicate all trefoils). With this convention, we had enough observations of the 7 nontrivial prime knot types with 6 or fewer crossings to perform statistical analyses. We made sporadic observations of~\emph{much} more complicated knots; for instance, the most complicated summands we observed were the 16-crossing knots $16a_{\num{61059}}$ and $16n_{\num{521027}}$. These knot types are clearly much rarer than the $\leq 6$ crossing knots, but our data on them is too sparse to quantify this precisely.

Both the raw data and summary statistics are archived on Dryad~\cite{dryad}. The summary statistics are also included as Supplemental Data.

\section{Results}

\begin{figure*}[ht]
\hfill
\begin{overpic}[width=0.45\textwidth]{{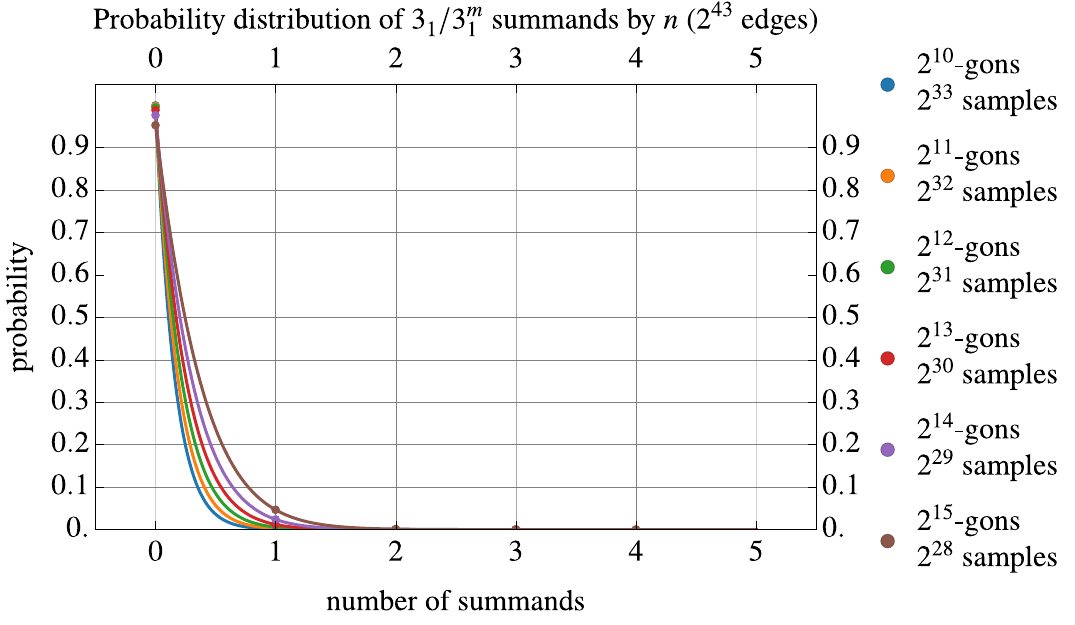}}
\end{overpic}
\hfill
\begin{overpic}[width=0.45\textwidth]{{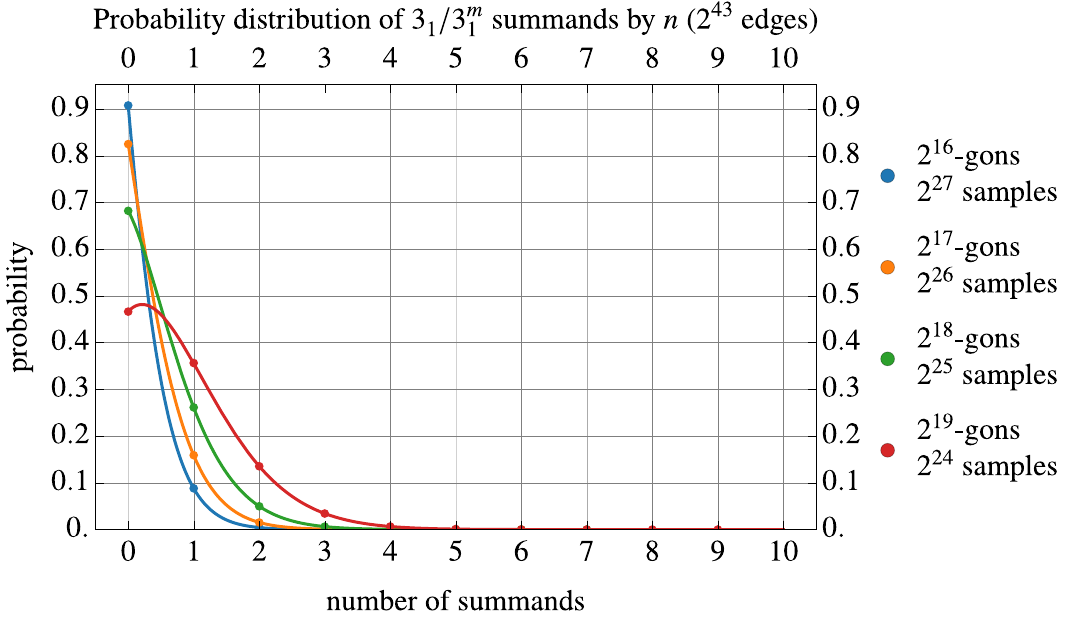}}
\end{overpic}
\hfill
\hphantom{.}
\\
\vspace{0.2in}
\hfill
\begin{overpic}[width=0.45\textwidth]{{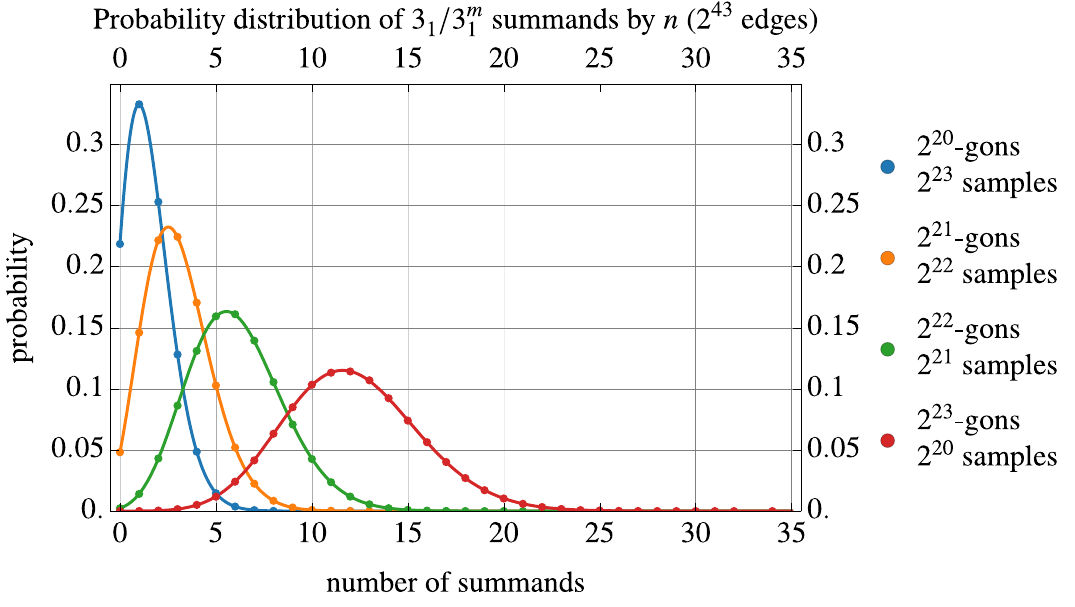
}}
\end{overpic}
\hfill
\begin{overpic}[width=0.45\textwidth]{{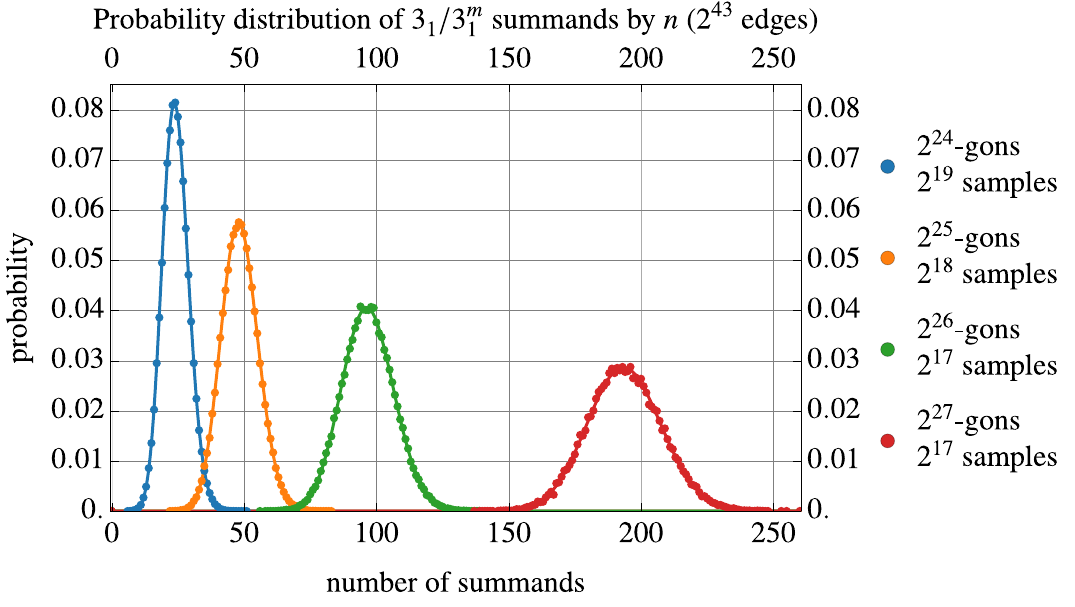}}
\end{overpic}
\hfill
\hphantom{.}

\caption{These plots show the probability $P(m_{3_1}(n)=m)$ of observing $m$ trefoil summands in off-lattice self-avoiding polygons of length $n$, together with the Poisson distribution function $\frac{(\lambda_K(n))^m e^{-\lambda_K(n)}}{m!}$ from~\eqref{eq:fundamental equation}. This is defined only at integer $m$, but we draw connecting curves as guides for the eye. The number of $n$-gons sampled ($2^{43}/n$) decreases with $n$, so the data appears rougher for large $n$.}
\label{fig:poisson}
\end{figure*}

\begin{figure*}[ht]
\hfill
\begin{overpic}[width=0.45\textwidth]{{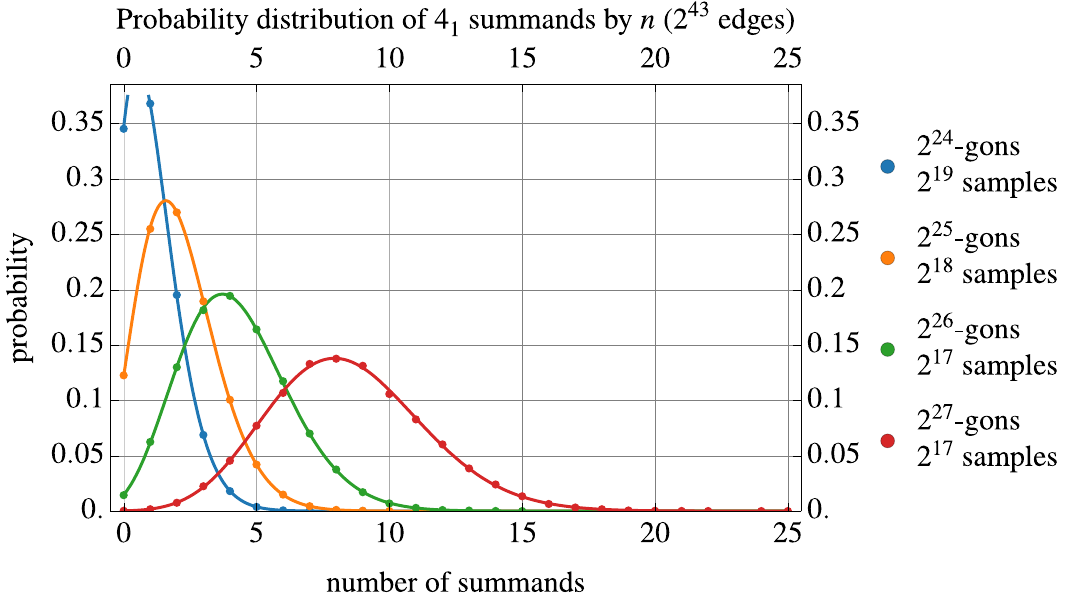}}
\end{overpic}
\hfill
\begin{overpic}[width=0.45\textwidth]{{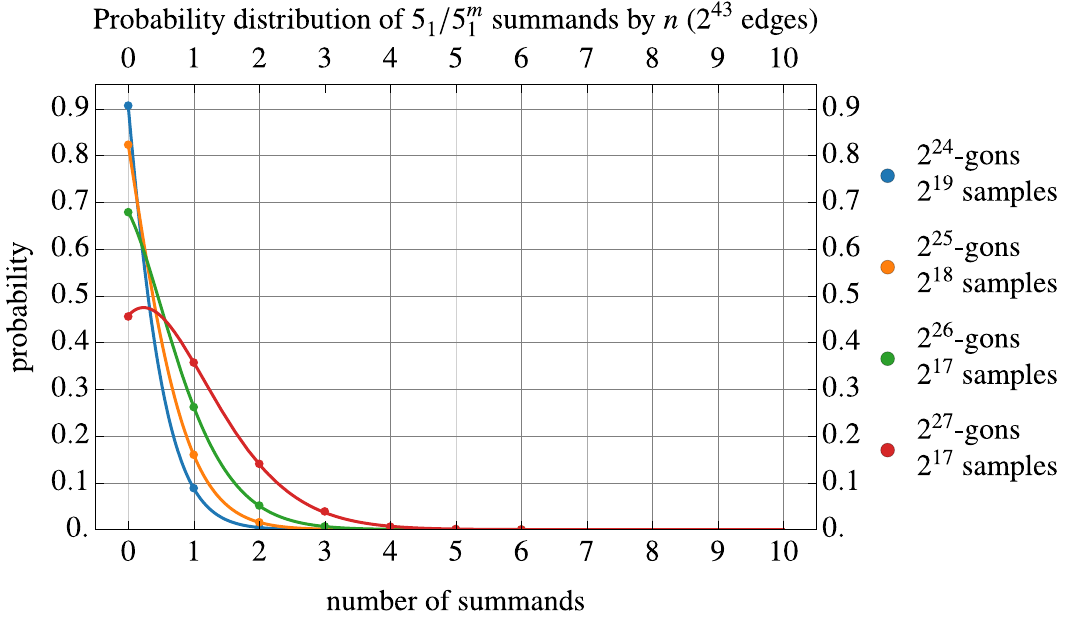}}
\end{overpic}
\hfill
\hphantom{.}
\\
\vspace{0.2in}
\hfill
\begin{overpic}[width=0.45\textwidth]{{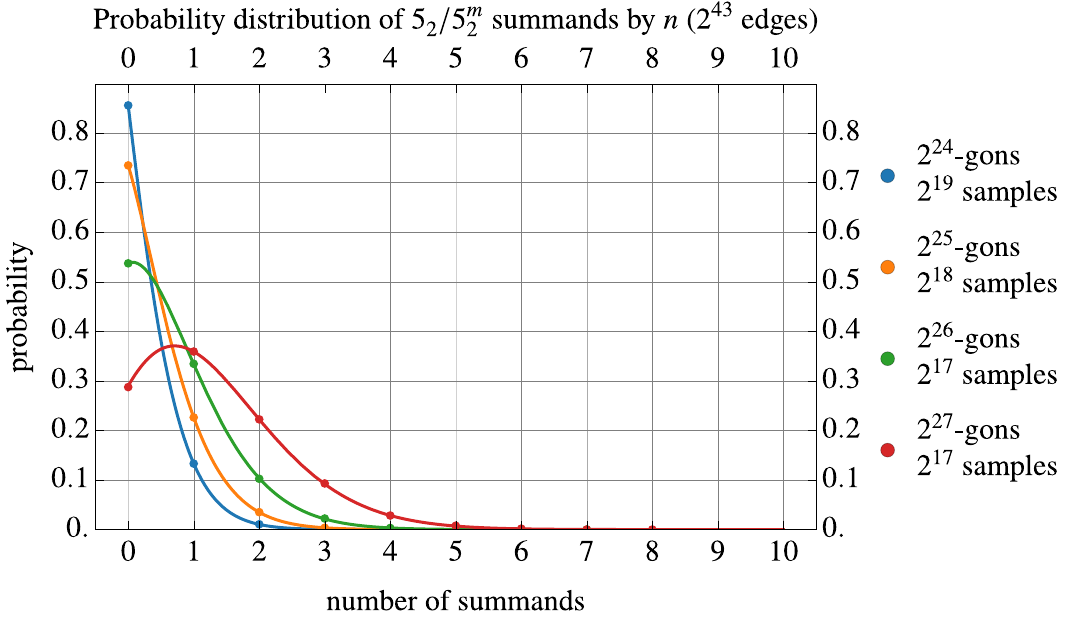}}
\end{overpic}
\hfill
\begin{overpic}[width=0.45\textwidth]{{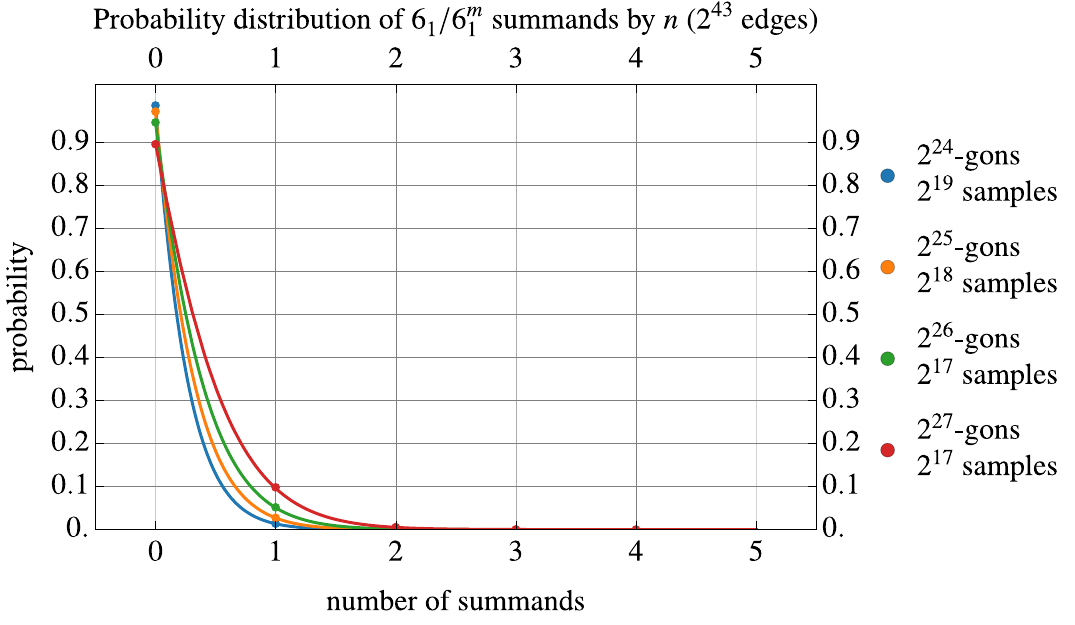}}
\end{overpic}
\hfill
\hphantom{.}

\caption{These plots show the probability $P(m_{4_1}(n)=m)$, $P(m_{5_1}(n)=m)$, $P(m_{5_2}(n)=m)$, and $P(m_{6_1}) = m)$ of observing $m$ summands of knot type $4_1$, $5_1/5_1^m$, $5_2/5_2^m$, or $6_1/6_1^m$ in off-lattice self-avoiding polygons of length $n$, together with the Poisson distribution functions $\frac{(\lambda_K(n))^m e^{-\lambda_K(n)}}{m!}$ from~\eqref{eq:fundamental equation}. Since these knots are much less probable than trefoils, only large values of $n$ are shown. The Poisson fit remains very good.}
\label{fig:poissontwo}
\end{figure*}

We restricted our analysis to $K$ and $n$ where we observed at least $10$ summands of type $K$ in each of the parallel Markov chains. This left us with the $7$ knot types with $\leq 6$ crossings. We used Geyer's IPS estimate~\cite{geyerPracticalMarkovChain1992} for the sample mean and standard error of mean for $m_K^n$ for each of the $\geq 64$ Markov chains, then combined these estimates across parallel chains (assuming independence). The result was an empirical value for $\lambda_K(n)$ with an error estimate. According to these estimates, our maximum relative (standard) error in $\lambda_K(n)$ was less than $3\%$. 

With these values of $\lambda_K(n)$, we were able to compute the Poisson distributions on the right-hand side of~\eqref{eq:fundamental equation}. \autoref{fig:poisson} and~\autoref{fig:poissontwo} show the empirical distribution of $m_K^n$ for trefoils, figure-8 knots, $5_1$, $5_2$, and $6_1$ for various $n$, together with the corresponding Poisson distributions.

\begin{figure}[ht]
	\centering
		\includegraphics[width=0.6\textwidth]{{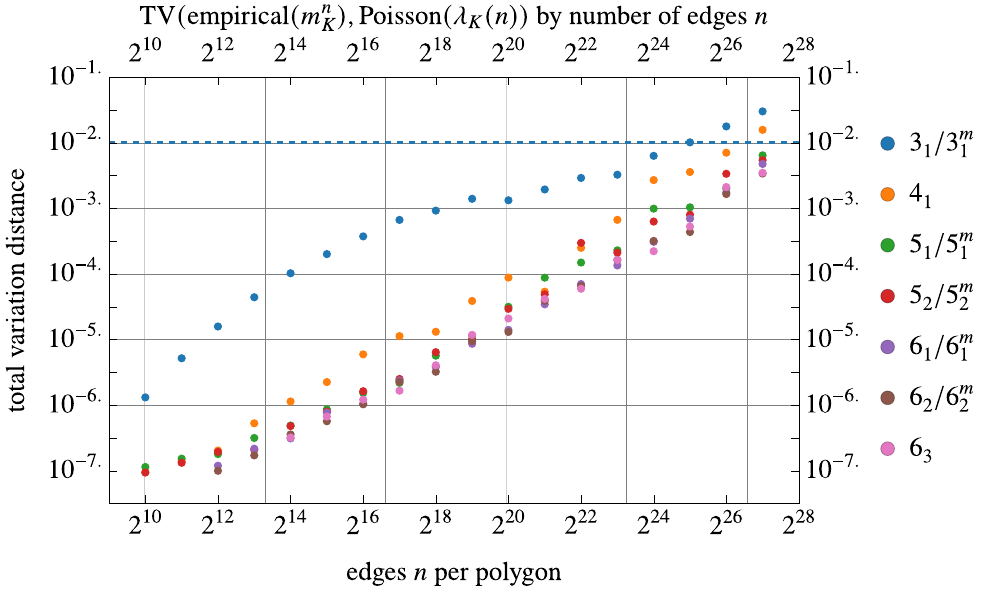}}
	\caption{This figure shows the total variation distance $\operatorname{TV}(p_1,p_2) = \sum_{m} |p_1(m) - p_2(m)|$ between the empirical distribution for the number of prime summands of knot type $K$ in an off-lattice self-avoiding $n$-gon $m_K^n$ and the corresponding Poisson model with mean $\lambda_K(n)$, using the estimate for $\lambda_K(n)$ from our dataset. We can see that almost all of these distances are very small, confirming that~\eqref{eq:fundamental equation} is an accurate approximation for these $K$ and $n$.}
	\label{fig:total variation}
\end{figure}

We computed the total variation distance (TV)~\footnote{This is the maximum difference between the probabilities of any event (such as $m_K^n = m$) according to the two distributions.} between the empirical distribution of $m_K^n$ and the Poisson model.  The results appear in~\autoref{fig:total variation}; they show that this distance is less than $0.01$ (and often much smaller) except for a few values of $m_{3_1/3_1^m}^n$ and $m_{4_1}^n$ where $n$ is very large. In these cases, $\operatorname{TV}$ was as large as $0.03$. Our estimates for $\lambda_K(n)$ with estimated errors and the total variation distances are given in Supplemental Data and in the Dryad dataset~\cite{dryad}.

At first, the rise in $\operatorname{TV}$ with $n$ might seem counterintuitive; the Poisson approximation should get better in the asymptotic limit. We think this an artifact of our experimental design, where the number of polygons sampled decreases with $n$. To check this, we performed Pearson's $\chi^2$ test to check the hypothesis that the large-$n$ data for $m_{3_1/3_1^m}^n$ and $m_{4_1}^n$ was sampled from the Poisson distribution. As expected, for $K = 3_1/3_1^m$ and $2^{23} \leq n \leq 2^{27}$ and $K = 4_1$ and $2^{18} \leq n \leq 2^{27}$ (covering all of the cases where $\operatorname{TV} \geq 0.01$), this test could not reject the hypothesis at $p = 0.05$.

\begin{figure*}[t]
\hfill
\begin{overpic}[height=0.25\textwidth]{{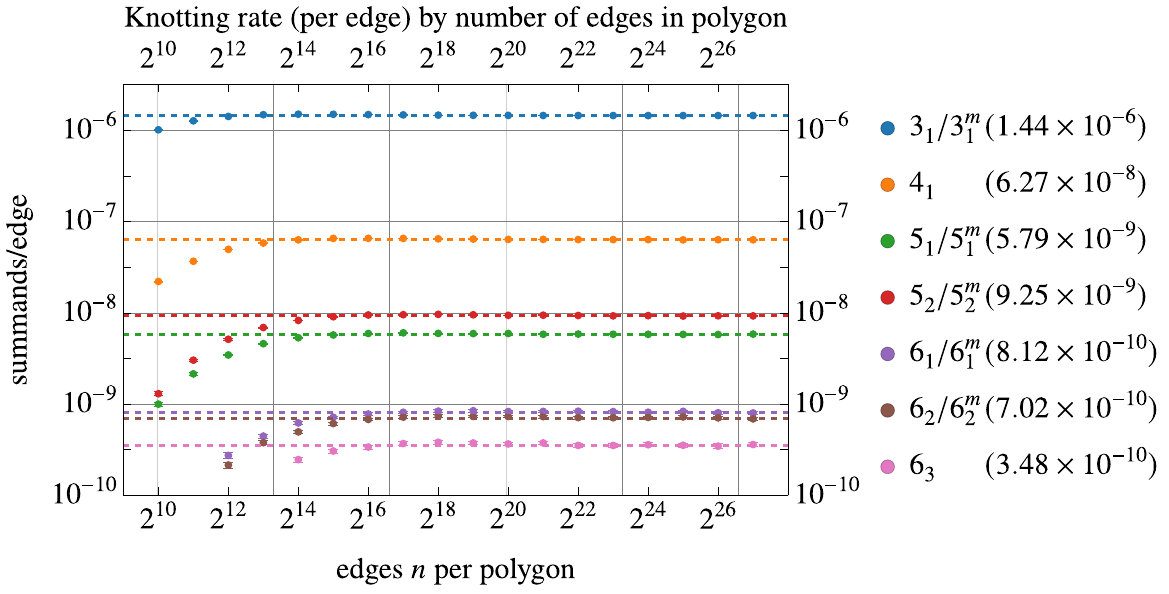}}
\end{overpic}
\hfill
\begin{overpic}[height=0.25\textwidth]{{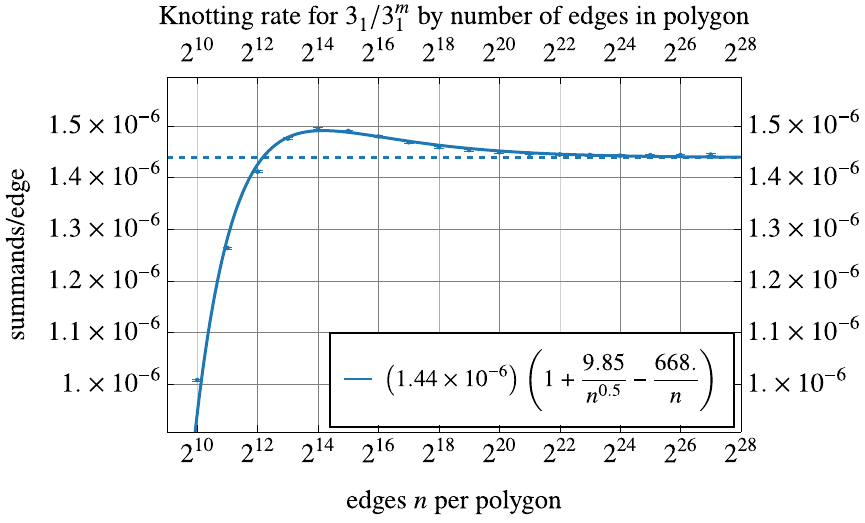}}
\end{overpic}
\hfill
\hphantom{.}

\caption{The log-log plot at left shows the rate of knotting per edge: $R_K(n) = \lambda_K(n)/n$ estimated by our experiment. $99\%$ confidence intervals are included, but are too small to be visible on the plot. The fact that these values are becoming constant in $n$ supports our conclusion that the asymptotic knotting rate $C_K = \lim_{n \rightarrow \infty} \lambda_K(n)/n$ exists. Our estimates for $C_K$ appear with their knot types. The semilog plot at right shows this probability for trefoils only, along with our proposed fit to the finite-size correction~\eqref{eq:rate with finite size effect model}.}
\label{fig:rates}
\end{figure*}

We then computed the summand production rate (per edge): $R_K(n) = \lambda_K(n)/n$ across our dataset. Again, we had enough data to compute these rates for knots with six and fewer crossings. \autoref{fig:rates} shows the result; all of the knotting rates have a finite-size correction, which differs by knot type. Further, \eqref{eq:rate with finite size effect model} with $\Delta = 0.5$ explains the finite-size corrections reasonably well. We notice that the asymptotic rate of production decreases very quickly as the knots become more complex, by more than an order of magnitude for each crossing number. This means that in this model the vast majority of knotting consists of trefoil and figure-8 summands, with all other knot types vanishingly rare in comparison. Baiesi, Orlandini, and Stella~\cite{BaiesiStella2010} and Rechnitzer and Janse van Rensburg~\cite{JANSEVANRENSBURG:2011jo} estimated the asymptotic ratios $C_K/C_K'$ by computing $P_K(n)$, finding that these ratios did not depend on the lattice. We can make corresponding estimates for our off-lattice model, computing $C_K$ from $R_K(n)$; the results appear in~\autoref{tab:amplitude ratios}.

\begin{table}[ht]
\centering
\begin{tabular}{l|ccc|ccc|ccc|cc|cc|cc}
-/-             & \multicolumn{3}{c}{$4_1$} & \multicolumn{3}{c}{$5_1/5_1^m$} & \multicolumn{3}{c}{$5_2/5_2^m$} & \multicolumn{2}{c}{$6_1/6_1^m$} & \multicolumn{2}{c}{$6_2/6_2^m$} & \multicolumn{2}{c}{$6_3$}  \\
& (new) & BOS & JvRR & (new) & BOS & JvRR & (new) & BOS & JvRR & (new) & BOS & (new) & BOS & (new) & BOS \\
\hline 
$3_1/3_1^m$ & 22.9 & 22.12 & 28 & 248 & 197 & 400 & 156 & 140 & 280 & 1770 & 2633 & 2050 & 1668 & 4140 & 2677 \\
$4_1$       & - & - & - & 10.8 & 8.89 & 15 & 6.78 & 6.33 & 9 &  77.2 & 119 & 89.4 & 75.4 & 180 & 121 \\
$5_1/5_1^m$ & - & - & - & - & - & - & 0.626 & 0.712 & 0.67 & 7.13 & 13.39 &  8.25 & 8.48 & 16.7 & 13.61  \\
$5_2/5_2^m$ & - & - & - & - & - & - & - & - & - & 11.4 & 18.08 & 13.2 & 11.9 & 26.6 & 19.12 \\
$6_1/6_1^m$ & - & - & - & - & - & - & - & - & - & - & - &  1.16 & 0.63 & 2.34 & 1.01 \\
$6_2/6_2^m$ & - & - & - & - & - & - & - & - & - & - & - & - & - & 2.02 & 1.60 
\end{tabular}
\caption{This table gives the amplitude ratios $C_K/C_{K'}$, where $K$ is constant on rows, and $K'$ is constant on columns. Our data appears in the columns marked (new), the columns marked BOS contain simple cubic lattice data for $n$ up to $\num{200000}$ from Baiesi, Orlandini, and Stella~\cite{BaiesiStella2010}. The columns marked JvRR contain data for $n$ up to $512$ on the SCC, FCC, and BCC lattices from Janse van Rensburg and Rechnitzer~\cite{JANSEVANRENSBURG:2011jo}. The fact that our data is comparable to these lattice studies supports the idea that the bead-chain and lattice models all belong to the same universality class.}
\label{tab:amplitude ratios}
\end{table}

\begin{figure}[t]
	\centering
		\includegraphics[width=0.6\textwidth]{{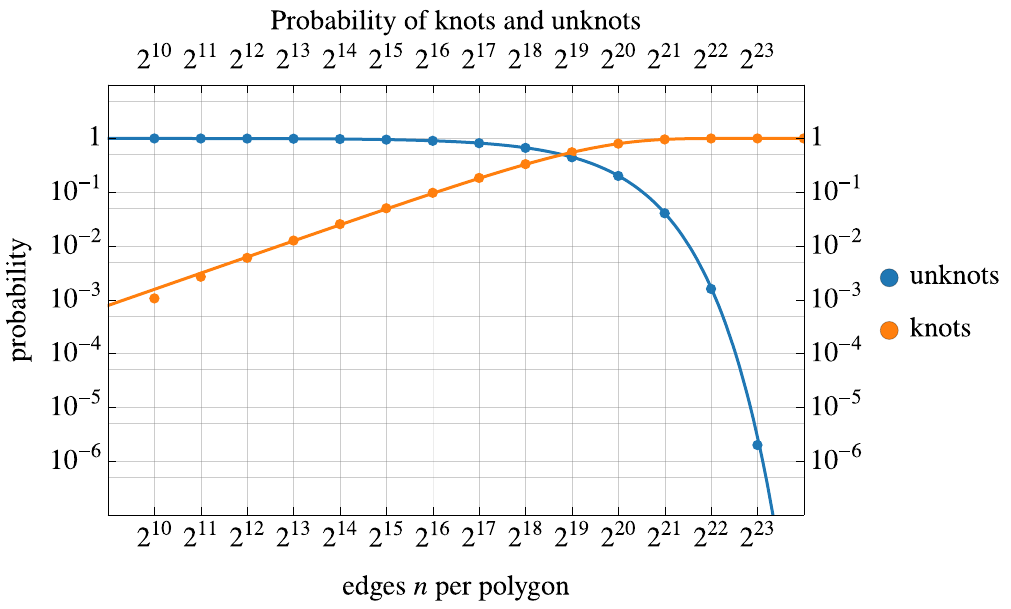}}
	\caption{Fraction of knots and unknots for $n$-edge SAPs, together with plots of $e^{-n/n_{0}}$ and $1-e^{-n/n_{0}}$ with the characteristic length $n_{0} = \num{656500}$ estimated from knotting rates. The plot stops at $n=2^{23}$ because we observed no unknotted SAPs for larger $n$.}
	\label{fig:unknot probability}
\end{figure}

We can directly measure knotting rates $R_K(n)$ for only a few knots. According to~\eqref{eq:unknot probability}, 
\[
\sum_{K} R_K(n) = -\frac{1}{n} \log P_{0_1}(n) 
\] 
where the sum is over all prime knot types $K$. Since we can measure $P_{0_1}(n)$ for small $n$ by counting unknotted polygons in our sample, we can use this to get a rough estimate of the production rate of more complex knots 
\[
\begin{aligned}
R_{\operatorname{complex}}(n) &:= \sum_{K\,:\, \operatorname{cr}(K) > 6} R_K(n) = -\frac{1}{n} \log P_{0_1} -\sum_{K\,:\, \operatorname{cr}(K) \leq 6} R_K(n) 
\end{aligned}
\]
Using this method, a conservative estimate is that for $2^{10} \leq n \leq 2^{21}$, we have $R_{\operatorname{complex}}(n) < 10^{-8}$. Assuming that this bound holds for larger $n$, we can estimate the characteristic length $n_0 = 1/\sum_K C_K$ by $n_0 \approx \num{656500} \pm \num{2500}$. \autoref{fig:unknot probability} shows that for large $n$, the approximation $P_{0_1}(n) \approx e^{-n/n_0}$ provides a quite useful estimate of the probability of unknotting and the probability of knotting.
We note that~\cite{KoniarisMuthukumar1991} estimates $n_0 \approx \num{800000}$ for this model based on simulations with $\leq \num{1000}$ beads. 

\begin{figure}[t]
\centering
		\includegraphics[width=0.6\textwidth]{{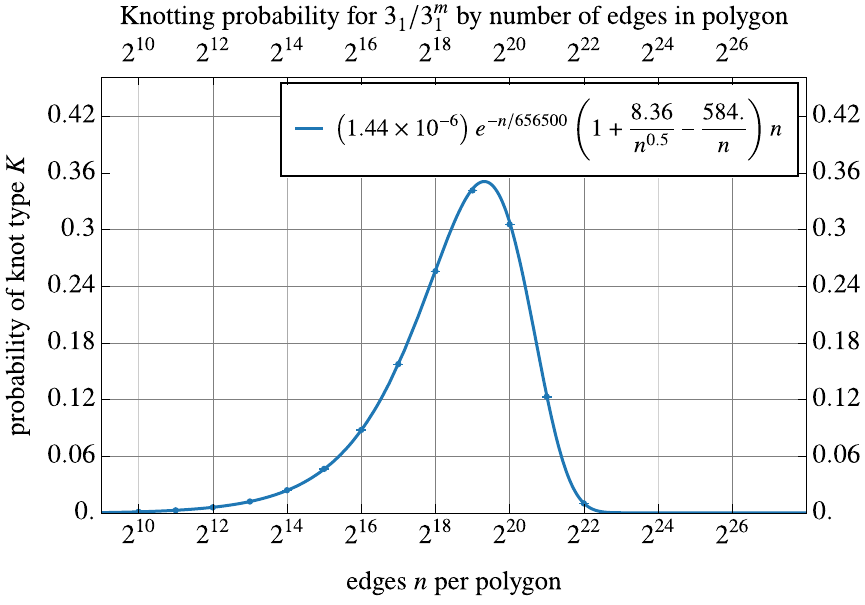}}
	\caption{The probability of knot type $K = 3_1$ or $K = 3_1^m$, along with $99\%$ confidence intervals, based on observed counts in our dataset (the errors were estimated with Geyer's IPS estimate for each Markov chain, then combined across parallel chains assuming independence). The data fits well to the standard model~\eqref{eq:knot probability with finite size effect model}. The coefficient $C_K$ and the $\beta_K$ and $\gamma_K$ for the finite-size correction are comparable to those measured for $R_K(n)$ shown in~\autoref{fig:rates}.}
	\label{fig:trefoil probability}
\end{figure}

We can now go back to the original knot entropy conjecture, and compare the observed probabilities $P_K(n)$ for various prime knot types with the predictions from~\eqref{eq:standard model}. Our goal is to check that the prediction of knot probabilities is good, and that the fitted values for $C_K$, $\beta_K$ and $\gamma_K$ are comparable to those we obtained by $R_K(n)$ by fitting~\eqref{eq:rate with finite size effect model}.~\autoref{fig:trefoil probability} shows the result for the trefoil knot. Here, where we have many observations of knots with knot type $3_1/3_1^m$, the results are all quite similar. For the remaining knot types, we have a lot less data for $P_K(n)$ and the results are different; fitting $R_K(n)$ predicts asymptotic values for $C_K$ which are about $10\%$ lower than the values obtained by fitting $P_K(n)$. The complete collection of results is shown in~\autoref{tab:main data table}. 

\begin{table*}[ht]
\centering
\begin{tabular}{l@{\hskip .5em}|@{\hskip 1em}ll@{\hskip 1em}ll@{\hskip 1em}ll@{\hskip 1em}ll}
 & $C_K$ & & $\beta_K$ & & $\gamma_K$ & & \multicolumn{2}{l}{adjusted $R^2$} \\ \hline
$3_{1}^{}/3_{1}^{m}$ & $1.44\times 10^{-6}$ & $(1.44\times 10^{-6})$ & $9.85$ & $(\phantom{-}8.36)$ & $-668.$ & $(-584.)$ & $1.$ & $(0.99991)$\\ 
$4_{1}^{}$ & $6.27\times 10^{-8}$ & $(6.71\times 10^{-8})$ & $19.4$ & $(-8.94)$ & $-2390.$ & $(-424.)$ & $0.99999$ & $(0.99837)$\\ 
$5_{1}^{}/5_{1}^{m}$ & $5.79\times 10^{-9}$ & $(6.53\times 10^{-9})$ & $17.1$ & $(-29.8)$ & $-3320.$ & $(\phantom{-}67.8)$ & $0.99992$ & $(0.99651)$\\ 
$5_{2}^{}/5_{2}^{m}$ & $9.25\times 10^{-9}$ & $(1.06\times 10^{-8})$ & $17.9$ & $(-35.3)$ & $-3780.$ & $(\phantom{-}219.)$ & $0.9999$ & $(0.99662)$\\ 
$6_{1}^{}/6_{1}^{m}$ & $8.12\times 10^{-10}$ & $(9.04\times 10^{-10})$ & $28.7$ & $(-37.1)$ & $-9370.$ & $(-520.)$ & $0.99981$ & $(0.99857)$\\ 
$6_{2}^{}/6_{2}^{m}$ & $7.02\times 10^{-10}$ & $(8.16\times 10^{-10})$ & $45.5$ & $(-48.1)$ & $-12800.$ & $(\phantom{-}44.8)$ & $0.9998$ & $(0.99862)$\\ 
$6_{3}^{}$ & $3.48\times 10^{-10}$ & $(3.89\times 10^{-10})$ & $86.2$ & $(-14.4)$ & $-22600.$ & $(-4150.)$ & $0.99962$ & $(0.99915)$
\end{tabular}
\caption{This table shows our final values for the amplitude $C_K$ and the finite-size correction coefficients $\beta_K$ and $\gamma_K$ computed by fitting the observed rate of knotting $R_K(n)$ from our dataset to~\eqref{eq:rate with finite size effect model}. The last column shows the adjusted $R^2$ value for the fit to $R_K(n)$ rate as a function of $n$. These fits are extremely good. In parentheses, we see the corresponding values predicted by fitting~\eqref{eq:knot probability with finite size effect model} to our much more limited data on $P_K(n)$, with the last column showing the adjusted $R^2$ for the fit to $P_K(n)$. For the trefoil, these agree quite well. In the remaining cases, we think that the $R_K(n)$ predictions are more accurate, as we have data for much larger $n$ in these cases.}
\label{tab:main data table}
\end{table*}

\section{Conclusion and Future Directions}

We have presented new evidence consistent with knot localization and the knot entropy conjecture for very large off-lattice self-avoiding polygons in the ``bead-necklace'' model. Three things made our experiments possible: switching our focus from computing the probability of individual knot types to counting summands in the prime decomposition, which gave us a new observable to target, an implementation of Clisby's tree for polygons, which gave us large polygons, and a new knot simplification and classification code, which allowed us to compute our observable.

Previous experiments have been essentially limited by the difficulty of computing knot invariants for large polygons. The limiting factor for our experiment turned out to be the time required to burn in the polygon-generating Markov chain, which consumed the majority of the computation time for $2^{26}$- and $2^{27}$-gons (cf.\ Section 4 of~\cite{ClisbyEfficient2010}). We don't think that the number of steps for burn-in can be reduced from the current $20\,n$. So it would be very interesting to find a better strategy for generating starting polygons.

The difference between the amplitudes $C_K$ computed by fitting $R_K$ and $P_K$ for non-trefoil knots is also an interesting issue. We think the difference is mostly a matter of data quality; a much larger experiment with many more polygons might observe enough ``pure'' figure-8 knots, for instance, to make the $C_K$ figures agree. However, it is also possible that the finite-size effect model needs to be refined. Again, we leave this to future work.

Whittington suggests~\cite{WhittingtonPC} that the knotting rate $R_K(n)$ should continue to increase with $n$, perhaps very slowly, as larger and larger knots of type $K$ occur. This is a very interesting idea. We do see a very slight rise in $R_{3_1/3_1^m}$ for $n=2^{26}$ and $n=2^{27}$, though our error bars do not allow us to resolve this unambiguously. Ultimately, this question will have to be resolved by a theorem, or by much larger experiments than we do here. 

Our knot simplification and identification toolchain~\cite{schumacherKnoodle} works well for much larger polygons, and is independent of the model used to generate the polygons. We therefore hope that other authors will attempt similar on-lattice experiments, in particular to try to refine the estimates of amplitude ratios on the lattice. It would also be interesting to apply our toolchain to non-self-avoiding polygons, and we intend to do so. Preliminary tests indicate that these have many simple summands, but that they also seem to have one or two prime factors with much more complicated knotting. 

Of course, now that we can generate very large example polygons containing many trefoil summands, it would be interesting to see if they can be geometrically isolated and their sizes measured directly. We leave this challenge to future work, as it is nontrivial, but it is encouraging that the Poisson model in~\eqref{eq:fundamental equation} is quite compatible with the knot localization conjecture that most prime factors should be tight and isolated from each other.

Finally, note that the ``sum rules'' of~\cite{UeharaKnotting2017,deguchiSumRules2020} for $P_{K \# K'}(n)$ follow as immediate consequences of the Poisson model in~\eqref{eq:fundamental equation}. They are also verified by the empirical measurements of $P_{K \# K'}$ in our dataset. 

\begin{acknowledgments}
We are very grateful to our colleagues for many helpful discussions, especially Nick Beaton, Mark Dennis, Shura Grosberg, Esias Janse van Rensburg, Enzo Orlandini, Andrew Rechnitzer, Chris Soteros, and Stuart Whittington. Several important improvements in this paper came from BIRS workshop 25w5490, \emph{Polymer Modeling and DNA Topology: The Interplay between Theory, Computation and Experiments} and some crucial work was done at BIRS during 25rit038, \emph{Stick Numbers and Polygonal Knot Theory}. We appreciate the support of the Japan Science and Technology Agency (CREST Grant Number JPMJCR19T4) and the National Science Foundation (DMS--2107700). H.S.~was supported by the Research Training Group ``Energy, Entropy, and Dissipative Dynamics (EDDy)'', funded by the Deutsche Forschungsgemeinschaft (DFG, German Research Foundation), project no.\ 320021702/GRK2326.
\end{acknowledgments}





\bibliography{specials.bib,zotero-library-current.bib,bib_TetsuoErica}










\end{document}